\documentclass[letterpaper,11pt]{article}
\usepackage{jheppub}
\usepackage[english]{babel}
\usepackage{amsmath,amssymb,graphicx,xcolor}
\usepackage[font=normalsize]{subfig}
\usepackage{slashed}
\usepackage{ulem}

\newcommand{\tmop}[1]{\ensuremath{\operatorname{#1}}}

\preprint{}

\allowdisplaybreaks

\title{Decay of the charged Higgs boson and the top quark in two-Higgs-doublet model at NNLO in QCD}

\author[a]{Xiao-Min Shen,}
\emailAdd{xmshen137@sjtu.edu.cn}

\author[a]{YaLu Hu,}
\emailAdd{018072910016@sjtu.edu.cn}

\author[a]{ChuanLe Sun,}
\emailAdd{chlsun60@sjtu.edu.cn}

\author[a,b,c]{Jun Gao}
\emailAdd{jung49@sjtu.edu.cn}

\affiliation[a]{INPAC, Shanghai Key Laboratory for Particle Physics and Cosmology,
School of Physics and Astronomy, Shanghai Jiao Tong University, Shanghai 200240, China}
\affiliation[b]{Key Laboratory for Particle Astrophysics and Cosmology, Shanghai 200240, China
}
\affiliation[c]{Center for High Energy Physics, Peking University, Beijing 100871, China}

\abstract{
  We present numerical calculations of the partial width of the
  charged Higgs boson decay into a top quark,
  $H^- \rightarrow \bar{t} + b + X$, and the partial width of the top
  quark decay into a light charged Higgs boson
  $t \rightarrow H^+ + b + X$ at next-to-next-to-leading order (NNLO)
  in QCD, based on a factorization formula of the jet mass.  The NNLO
  corrections significantly reduce the renormalization scale
  dependence of the partial decay width in both cases.  We show
  relative size of the NNLO corrections for different charged Higgs
  boson masses and for different renormalization scales.  The NNLO
  corrections are about 16\% (1\%) of the leading order widths
  for the charged Higgs boson mass of
  200~GeV (2000~GeV), while it is quite small for the top quark
    decay.  Our analyses are independent of the detailed structure of
  the Yukawa couplings, and can be applied to various new physics
  models, as demonstrated by the decay branching ratio in different
  types of the two-Higgs-doublet models.

}


\begin{document}
\maketitle
\newpage

\section{Introduction}
The discovery of the Higgs boson by the ATLAS and CMS
experiments {\cite{1207.7214,1207.7235}} at the Large Hadron Collider (LHC)
makes a milestone in particle physics. Further measurements at the LHC have shown that
properties of the 125 GeV Higgs boson are consistent with the predictions of
the standard model (SM) up to current accuracy. Nevertheless, there
are numerous motivations suggesting that the scalar sector is likely to be
non-minimal. For example, the existence of extended Higgs sectors may help to
explain origins of the neutrino masses, dark matter, and matter-antimatter asymmetry.
Extra Higgs sectors are also needed in such as the supersymmetric
models, the Peccei-Quinn model {\cite{PhysRevLett.38.1440}} etc. An interesting
{feature} of the extended Higgs sectors is the possibility of one or
more charged scalars (also known as the charged Higgs boson), which is the
topic of this study. One of the simplest models that contain an extra Higgs
sector is the two-Higgs-doublet models (2HDM)
{\cite{PhysRev.D8.1226, 1106.0034}}. In a 2HDM, there are three neutral
scalars (one of them is a pseudo-scalar if CP is conserved) and two charged
scalars $H^{\pm}$. The study on properties of the charged Higgs boson can be
essential to distinguish different types of 2HDM {\cite{1106.0034, cheung2022comprehensive}}.

In this work, we refer to { the} charged Higgs boson with mass $m_H$ larger/less than the
top quark mass as heavy/light charged Higgs boson. Due to the large couplings between the
charged Higgs boson and the heavy fermions in new physics models, such as the 2HDM, one
of the promising production channel at the LHC for heavy charged Higgs boson is $g g
\rightarrow \bar{t} b H^+$, while light charged Higgs boson can be produced via top
quark pair production with one of the top quark decaying into a charged Higgs boson.
Both the ATLAS and CMS collaborations have searched for the
charged Higgs boson, and these measurements can be classified by how the resonant
charged Higgs boson decays, such as
the $H^{\pm} \rightarrow t b$ channel \cite{1808.03599, 2102.10076, 1908.09206,2001.07763},
the $H^{\pm} \rightarrow c s / c b$ channel \cite{ATLAS-CONF-2021-037, 2005.08900, He:1998ie, Balazs:1998sb, Diaz-Cruz:2001igs},
the $H^{\pm} \rightarrow \tau \nu$ channel \cite{1807.07915,1903.04560},
or the bosonic decay channels \cite{1905.07453, ATLAS-CONF-2021-047, 1806.01532, 2104.04762}.
The couplings and mass of the charged Higgs boson in specific new physics
models can also be constrained indirectly by e.g., $\bar{B} \rightarrow X_s
\gamma$, $B \rightarrow \tau \nu$ {\cite{0706.2997}}.

On the side of theoretical calculations, the next-to-leading order (NLO) QCD corrections to the
decay widths of both the top quark $t \rightarrow H^+ + b +
X$ and the charged Higgs boson $H^- \rightarrow \bar{t} + b + X$ in the limit of $m_b \rightarrow 0$ have been known for a long time~{\cite{hep-ph/9208240, hep-ph/9301237,PhysLett.B252.461, PhysRev.D43.855, PhysLett.B269.155,
PhysRev.D51.218}}.
In this paper, we are dedicated to the numerical calculations of the partial decay width of the heavy
charged Higgs boson $H^- \rightarrow \bar{t} + b + X$, and of the top
quark decaying into a light charged Higgs boson
$t \rightarrow H^++ b + X$ { at next-to-next-to-leading order (NNLO) in QCD}, using a phase space slicing method~\cite{1210.2808}.

In recent years, there have been enormous advances in the higher-order { QCD}
calculations.  As to the higher-order corrections to the decay of
scalar bosons, the partial width of $H \to b\bar{b}$ is known up to
the next-to-next-to-next-to-next-to-leading order (N$^4$LO), in
the limit where the mass of the bottom quark is
neglected~\cite{Baikov:2005rw,Davies:2017xsp,Herzog:2017dtz}.
The partial width for $H \to gg$ has been calculated to the N$^3$LO~\cite{Baikov:2006ch}
and N$^4$LO~\cite{Herzog:2017dtz} in the heavy top-quark limit.
We refer the readers to~\cite{Denner:2011mq, Spira:2016ztx} for a complete
list of relevant calculations.
The fully differential cross sections
for $H \to b\bar{b}$ have been calculated to NNLO
in~\cite{Anastasiou:2011qx, DelDuca:2015zqa}
and N$^3$LO in~\cite{Mondini:2019gid} for massless bottom quarks, and to
NNLO in~\cite{Bernreuther:2018ynm,1907.05398,1911.11524,2007.15015}
with massive bottom quarks.
On the other hand, there is a long history for calculation for
higher-order corrections to heavy quark decays.  The NLO QCD corrections
to the top quark decay width were calculated in
\cite{Jezabek:1988iv,Czarnecki:1990kv,Li:1990qf}.
The NNLO corrections were calculated in the large top quark mass limit in Refs.
\cite{hep-ph/9806244, hep-ph/9906273, hep-ph/0403221}, and the full
NNLO corrections were given in \cite{1210.2808, 1301.7133}.
The top quark decay width via model-independent flavor-changing neutral current couplings was
calculated to NLO in QCD \cite{0810.3889, 1004.0898}.
The QCD corrections to the decay $b \to c \ell \nu$ have been known to
N$^3$LO  \cite{2011.13654}.
Very recently there have been several implementations towards matching hadronic decays of
the Higgs boson at NNLO with parton shower.
Ref.~\cite{1912.09982} presents the matched results for the Higgs boson decaying into
{massless bottom quarks within POWHEG framework~\cite{1002.2581} by the MiNLO method~\cite{1206.3572}.}
Ref.~\cite{2009.13533} {presents the calculation for Higgs decays} to massless bottom quarks as well as to
gluons within the GENEVA framework~\cite{1211.7049}.
Furthermore, in Ref.~\cite{Hu:2021rkt} the matching on decays has been extended to including
massive bottom quarks by merging of samples with different jet multiplicities.

The rest of our paper is organized as follows.
In section \ref{sec:framework}, we present the framework { of} our fixed-order calculations.
Section \ref{sec:num-res} gives { the} numerical results including
NNLO partial decay width for $H^- \rightarrow \bar{t} + b + X$ and $t \rightarrow H^++ b + X$
for different charged Higgs boson masses, together with applications to benchmark scenarios in type-II and type-X
2HDM. We conclude in section \ref{sec:summary}.
\section{Framework}\label{sec:framework}

\subsection{Effective operator} \label{sec:eff-oper}
The interactions between the charged Higgs boson and quarks can be expressed
as
\begin{eqnarray}
  \mathcal{L}_{H^{\pm}} & = & \sum_{i, j} \bar{u}_i (Y^{(1)}_{i j} P_L +
  Y^{(2)}_{i j} P_R) d_j H^+ + \text{h.c.} \,,
\end{eqnarray}
where $u_i \in \{ u, c, t \}$ and $d_j \in \{ d, s, b \}$ are up-type and
down-type quarks, { respectively.} $P_L( P_R)$ is the left(right)-handed projection
operators, and $Y^{(1,2)}_{i j}$ are { the corresponding complex matrix elements of Yukawa
couplings.}

{ Potential applications of this effective operator  in phenomenology can be found in the studies of}  two-Higgs-doublet
models (2HDM) without tree-level flavor-changing neutral currents
\cite{1106.0034}. For example, in a type-II 2HDM, the corresponding Yukawa
couplings read
\begin{eqnarray}
  Y^{(1)}_{i j} & = & \frac{\sqrt{2}}{v} V_{u_i d_j} m_{u_i} \cot \beta\,,
  \nonumber\\
  Y^{(2)}_{i j} & = & - \frac{\sqrt{2}}{v} V_{u_i d_j} m_{d_j} \tan \beta\,,
  \nonumber
\end{eqnarray}
where $V$'s are CKM matrix elements, $v$ is the vacuum expectation value,
$\beta$ is the rotation angle which diagonalizes the mass-squared matrices of
the charged Higgs bosons and of the pseudo-scalars.

In this work, we focus on the interaction between charged Higgs bosons and
third-generation quarks. We assume the bottom quark to be massless except for
{ its non-vanishing} Yukawa coupling\footnote{For example, in type-II 2HDM, the $m_b
\tan \beta$ term may be competitive with or even larger than the $m_t \cot
\beta$ term if $\tan \beta \gg 1$, which is the region that experimental results
prefer.}, { then the cross section is proportional to
$| Y^{(1)}_{33} |^2 + | Y_{33}^{(2)} |^2$.}
So our calculation is essentially independent of the details of the Yukawa
coupling, and our results can be applied to different types of
two-Higgs-doublet models. Note, however, the renormalization of the Yukawa
coupling matters in our work, which can be found in Appendix~\ref{app:FO}. Since the
detailed structure of $Y^{(i)}$ is {beyond the scope of our paper}, we will mainly focus on the
ratio of the NNLO corrections to that of the LO contribution (as known as the
$K$ factors).
For both LO and NNLO contributions given in our work, the Yukawa couplings run at
three-loop and match at two-loop near thresholds, such
that the Yukawa couplings are canceled out in these $K$ factors. 
As a result, the $K$ factors given in our work will be independent of
the details of the Yukawa coupling, and may be applied to different
types of two-Higgs-doublet models.
\subsection{Phase space slicing method}

Our calculation of the NNLO decay width of charged Higgs boson or top quark is
based on the phase space slicing method \cite{1210.2808}.
For both top quark decay $t \rightarrow H^+ +
b + X$ and charged Higgs boson decay $H^- \rightarrow \bar{t} + b + X$, where $X$
are massless partons or bottom quarks, we cluster all the partons
in the final state into a single jet, the mass of which is defined as
\begin{eqnarray}
  m_J^2 & {\equiv} & (p_t - p_H)^2 \,. \label{eq:jet-mass}
\end{eqnarray}
According to the cutoff parameter
\begin{eqnarray}
  \rho & \equiv & \frac{m_J m_t}{| m_H^2 - m_t^2 |},  \label{eq:rho}
\end{eqnarray}
{the phase space can be divided} into two regions, the resolved part where $\rho
> \rho_{\tmop{cut}}$, and the unresolved part where $\rho \leqslant
\rho_{\tmop{cut}}$ {,  with $\rho_{\tmop{cut}}\ll 1$}. {Then the NNLO partial
decay width can be rewritten as,}

\begin{eqnarray}
  \Gamma & = & \int_0^{\rho_{\tmop{cut}}} \frac{d \Gamma}{d \rho'} d \rho' +
  \int_{\rho_{\tmop{cut}}}^{\rho_{\max}} \frac{d \Gamma}{d \rho'} d \rho'
  \nonumber\\
  & \equiv & \Gamma_{\tmop{unres}} + \Gamma_{\tmop{res}}  \,. \label{eq:slicing}
\end{eqnarray}

In the unresolved region, the contribution can be obtained
approximately at ${\cal O}(\rho^0_{\tmop{cut}})$ by factorization in
soft-collinear effective theory (SCET), and in the resolved region it
is calculated up to NNLO by Monte-Carlo simulation. The details are
given in section \ref{sec:FO}.

\subsection{Fixed-order calculation}\label{sec:FO}

Firstly, let us consider the {evaluation of decay width in the resolved region 
(including three-body and four-body phase spaces)}, 
taking heavy charged Higgs
decay { $H^- \rightarrow \bar{t} + b + X$} as an example. The NLO corrections to
$\Gamma_{\tmop{res}}$ are given by the tree level contribution of $H^-
\rightarrow \bar{t} + b + g$, which has no divergence in the $\rho >
\rho_{\tmop{cut}}$ region. The NNLO corrections to $\Gamma_{\tmop{res}}$
consist of two parts, the NLO corrections of $H^- \rightarrow \bar{t} +  b + g$,
denoted by $\Gamma_{t + 2 j}^{(2)}$, and the LO contribution $\Gamma_{t + 3
j}^{(1)}$ of $H^- \rightarrow \bar{t} + b + j + j$, where $j j$ may take $g g$, $q
\bar{q}$ or $b \bar{b}$.

Note that both $\Gamma_{t + 2 j}^{(2)}$ and $\Gamma_{t + 3 j}^{(1)}$ contain infrared (IR) divergences. 
The key point is that for $H^- \rightarrow \bar{t} + b + j + j$, the cut
$\rho > \rho_{\tmop{cut}}$ in four-body phase space forbids the appearance of
double unresolved partons, that is, there is at least one resolved parton. 
{ So we may regard the sum of $\Gamma_{t + 2 j}^{(2)}$ and $\Gamma_{t + 3 j}^{(1)}$ as the NLO
corrections to $H^- \rightarrow \bar{t} + b + g$, which are IR safe. The IR divergences of the two ingredients
can be removed individually by introducing appropriate dipole subtraction terms 
\cite{hep-ph/9602277, hep-ph/0201036, hep-ph/0408154, 1111.4991}, then they can be calculated numerically
by Monte-Carlo event generators. }

We neglect the masses of light quarks. The mass of bottom quark is also
omitted except for {its non-vanishing } Yukawa coupling. 
{The external gluon and quark fields are renormalized with on-shell (OS) scheme.}
The Yukawa coupling is renormalized in the $\overline{\rm MS}$ scheme. The
renormalization of QCD coupling is carried out in 5-flavor $\overline{\rm MS}$ 
scheme, with $\alpha_s^{(N_l = 5)} (m_Z) = 0.$1181. For completeness we
present ingredients of fixed-order calculation in Appendix~\ref{app:FO}. The
NLO amplitudes of three-body decays is generated by \textsc{FeynArts}
\cite{hep-ph/0012260}. They are further simplified and reduced to scalar
integral in the Passarino-Veltman reduction scheme \cite{PV-reduction} by \textsc{FeynCalc} \cite{1601.01167}.

The NLO decay widths in QCD for both top quark decay $t \rightarrow H^+ + b +
X$ \cite{hep-ph/9208240, hep-ph/9301237} and charged Higgs decay $H^- \rightarrow \bar{t} + b + X$
\cite{PhysLett.B252.461, PhysRev.D43.855, PhysLett.B269.155,
PhysRev.D51.218} in the $m_b \rightarrow 0$ limit have
been known for a long time , and we present them here for completeness.
{
\begin{eqnarray}
  \Gamma^{\tmop{NLO}}_{t \rightarrow H^+ b X} & = & \frac{1}{32 \pi} m_t \,
   y^2_{\tmop{MS}}   \left( 1 - \frac{m_H^2}{m_t^2} \right)^2
  \left[ 1 + \frac{16}{3} a_s \Delta_t + 2 a_s C_F \left( 3 \ln
  \frac{\mu^2}{m_t^2} + 4 \right) \right]  \label{eq:NLO-top}\,, \\
  \Delta_t   & = & - 2 \tmop{Li}_2
  (z) + \frac{z}{z - 1} \ln z + \left( \frac{1}{z} - \frac{5}{2} \right) \ln
  (1 - z) - \ln z \ln (1 - z) - 2 \zeta_2 + \frac{9}{4} \,, \nonumber\\
  &  &  \nonumber\\
  \Gamma^{\tmop{NLO}}_{H^- \rightarrow \bar{t} b X} & = & \frac{N_c}{16 \pi}
  m_H \,  y_{\tmop{MS}}^2   \left( 1 - \frac{m_t^2}{m_H^2} \right)^2
  \left[ 1 + \frac{16}{3} a_s \Delta_H + 2 a_s C_F \left( 3 \ln
  \frac{\mu^2}{m_t^2} + 4 \right) \right] \,, \label{eq:NLO-higgs}\\
  \Delta_H  & = & 2 \tmop{Li}_2
  (z) + \ln z \ln (1 - z) + \left( z - \frac{5}{2} \right) \ln \frac{1 - z}{z}
  + \frac{1}{z - 1} \ln z + \frac{9}{4} \,,\nonumber
\end{eqnarray}
where $z= m_H^2/m_t^2$ and $m_t^2/m_H^2$ for the top and charged Higgs decays, respectively.
$m_H$ is the mass of charged Higgs boson, $N_c = 3$ and $C_F = 4 /
3$ are color factors, $y^2_{\tmop{MS}}$ denotes the renormalized $| Y^{(1)}_{33} |^2 + |
Y_{33}^{(2)} |^2$ in the $\overline{\rm MS}$ scheme, and $a_s = \frac{\alpha_s
(\mu)}{4 \pi} = \frac{g_s^2}{(4 \pi)^2}$ is the strong coupling constant at
renormalization scale $\mu$, $\zeta_2 = \frac{\pi^2}{6}$. $\tmop{Li}_2$ is
polylogarithm of order 2.
}

\subsection{QCD factorization and singular distribution}\label{sec:singular-dist}

As is mentioned above, the decay width in the unresolved region
$\Gamma_{\tmop{unres}}$ defined in Eq.~(\ref{eq:slicing}) is calculated with
the help of the factorization formula in the threshold limit $m^2_J
\rightarrow 0$
\begin{eqnarray}
  \frac{1}{\Gamma_i^{\tmop{LO}}} \frac{d \Gamma_i}{d m_J^2} & = & H_i (m_H,
  m_t, \mu)  \int d p_J^2 d \omega J (p_J^2, \mu) S_i (\omega, \mu)
  \delta (m_J^2 - p_J^2 - 2 E_J^i \omega)  \,, \label{eq:factorization}
\end{eqnarray}
where the subscript $i$ takes $t, H$ for top quark decay $t \rightarrow H^+ + b + X$
and charged Higgs decay $H^- \rightarrow \bar{t} + b + X$ respectively.
$\Gamma_i^{\tmop{LO}}$, $H_i$, $S_i$ are the corresponding LO partial decay
width, hard function and soft function, { respectively}. The jet energy $E_J^i$ in the
threshold limit is given by
\begin{eqnarray}
  E_J^t & = & \frac{m_t^2 - m_H^2}{2 m_t} ~ \text{for top quark decay, } 
  \nonumber\\
  E_J^H & = & \frac{m_H^2 - m_t^2}{2 m_H} ~ \text{for charged Higgs decay. } 
\end{eqnarray}
The factorization formula Eq.~(\ref{eq:factorization}) is valid up to the
leading power in power expansion of $\rho_{\tmop{cut}}$ \cite{Boughezal:2016zws, Moult:2018jjd, Liu:2020tzd}. So the cut-off
$\rho_{\tmop{cut}}$ should be small enough such that the power corrections may
be safely omitted for phenomenological applications. In this work, an
empirical choice $\rho_{\tmop{cut}} = 3 \tmop{GeV} / m_H$ is used.
{The dependence of the decay width on $\rho_{\tmop{cut}}$ is discussed in section \ref{sec:num-res}}.
The heavy-to-light soft functions in Eq.~(\ref{eq:factorization}) for top
quark decay and charged Higgs decay, denoted by $S_t$ and $S_H$ respectively,
read
\begin{eqnarray}
  S_t(\omega) & = & \frac{1}{N_c} \sum_X {\rm Tr}
    \langle 0 | 
    \bar{Y}_v^{\dagger} (0) Y_n (0) |X\rangle
    \langle X | %\mathbf{T}
    Y_n^{\dagger} (0) \bar{Y}_v (0) |0\rangle
    \delta (\omega - n \cdot p_{\!_X}) \,,\nonumber\\
  S_H(\omega)& = & \frac{1}{N_c} \sum_X {\rm Tr}
    \langle 0 | Y_v^{\dagger} (0) Y_n (0) |X\rangle
   \langle X | %\mathbf{T}
     Y_n^{\dagger} (0) Y_v (0) |0\rangle
  \delta (\omega - n \cdot p_{\!_X})  \,, \label{eq:soft-def}
\end{eqnarray}
where the Tr is trace over color indices,
$n^\mu$ is the light-like vector in the jet direction.
Both quark jet function and soft functions have been known up to three loops
\cite{hep-ph/0603140, 1804.09722, hep-ph/0512208, 1911.04494}. More details about
the quark jet function and the soft functions are presented in Appendix~\ref{app::factorization}.
The hard function $H_i$ is the square of the Wilson coefficient of the $H^+
\bar{t} b$ operator, determined by matching from QCD to  SCET. In practice, they are 
  the virtual corrections to on-shell amplitude squares.
The NNLO Wilson coefficient
for charged Higgs decay is the same as the form factor for Goldstone
boson `decay', which can be found in \cite{0809.4687}. 
Because the $\mathcal{O} (\epsilon^2)$ terms at NLO 
($\epsilon=\frac{4-d}{2}$ is the dimensional regulator) 
is not present in \cite{0809.4687}, Ward identity is applied to extract the hard function of 
$t \rightarrow \bar{b} + H^+$ from the two-loop virtual corrections to $b
\rightarrow u$ decay~\cite{0810.1230}. 
The obtained hard function is expanded in
$\alpha_s^{(6)}$, which is then expressed in terms of 
$a_s^{(5)} \equiv \alpha_s^{(5)} / (4 \pi)$ as
\begin{eqnarray}
  H & = & 1 + a_s^{(5)} (\mu) H_1 + (a_s^{(5)})^2 H_2 +\mathcal{O}
  ((a_s^{(5)})^3) \,,
\end{eqnarray}
with the help of the decoupling relation
\begin{equation}
  \alpha_s^{(N_l + 1)} = \alpha_s^{(N_l)}  \left[ 1 + \frac{8}{3} T_F 
  \frac{\alpha_s^{(N_l)}}{4 \pi}  \left( - \frac{1}{2} L_\mu + \epsilon \left(
  \frac{L_{\mu}^2}{4} + \frac{1}{24} \pi^2 \right) \right) \right] \,,
  \label{eq:alpha-dec}
\end{equation}
where $N_l = 5$ is the number of light quark flavor, $T_F = 1 / 2$ for $\mbox{SU}(3)_c$, and
$L_{\mu} = \ln (m_t^2 / \mu^2)$.

Combining all these ingredients, {NNLO $\rho$ distribution at leading power is given by} 
\begin{eqnarray}
  \frac{1}{\Gamma_0} \frac{d \Gamma}{d \rho} & = & \delta (\rho) + a_s C_F \Big\{ (4
  L^2 + 10 L + 7 - 7 \zeta_2 + H_1 / C_F) \delta (\rho) - 14 L_0 (\rho) - 16
  L_1 (\rho)\Big\} 
 \nonumber\\ 
  &+& a_s^2 C_A C_F  \bigg\{ \frac{1}{648} \delta (\rho) \Big[6336 L^3 - 144 (36
        \zeta_2 - 299) L^2 - 24 (1008 \zeta_2 + 1188 \zeta_3 - 2935) L 
 \nonumber\\ &&
  + 50521 -45324 \zeta_2  - 22536 \zeta_3 + 16200 \zeta_4 \Big] + \frac{1}{9} L_0 (\rho)  (-
    924 L - 905 + 408 \zeta_2 + 72 \zeta_3)
 \nonumber\\ &&
  + \frac{4}{9} L_1 (\rho)  (- 264 L +
  95 + 72 \zeta_2) + 176 L_2 (\rho) \bigg\} 
 \nonumber\\
  &+&  a_s^2 C_F^2  \bigg\{ \delta (\rho) \Big[ 8 L^4 + 40 L^3 
   + \frac{1}{4}(312 - 112 \zeta_2) L^2 
   + \frac{1}{2}  (- 188 \zeta_2 + 96 \zeta_3 + 146) L
 \nonumber\\ &&
   + \frac{205}{8} - 94 \zeta_2 + 22 \zeta_3 + \frac{401 \zeta_4}{4} \Big] 
   +  L_0 (\rho)  (- 56 L^2 - 140 L - 101 + 66 \zeta_2 - 16 \zeta_3) 
  \nonumber\\ &&
   + 4 L_1 (\rho) (- 16 L^2 - 40 L + 21 + 12 \zeta_2) + 336 L_2 (\rho) + 128 L_3 (\rho)
  \bigg\} \nonumber\\
  & + & a_s^2 C_F N_l  \bigg\{ \frac{1}{324} \delta (\rho) \Big[ - 576 L^3 - 3600
        L^2 + 24 (72 \zeta_2 - 251) L - 4073 + 2628 \zeta_2 - 72 \zeta_3\Big]
  \nonumber\\ && -
  \frac{2}{9} L_0 (\rho)  (- 84 L - 85 + 24 \zeta_2) + \frac{8}{9} (24 L - 13)
  L_1 (\rho) - 32 L_2 (\rho) \bigg\} \nonumber\\
  & + & a_s^2 H_2 \delta (\rho) + a_s^2 C_F H_1 \big[ (4 L^2 + 10 L + 7 - 7
  \zeta_2) \delta (\rho) - 14 L_0 (\rho) - 16 L_1 (\rho) \big] \,,
  \label{eq:singular-explicit}
\end{eqnarray}
where $L  = \ln \frac{\mu m_t}{| m_H^2 - m_t^2|}$,
$L_n$ are plus distributions defined as
\begin{eqnarray}
  L_n & \equiv & \left[ \frac{\ln^n \rho}{\rho} \right]_+, n \geqslant 0 \,.
  \nonumber
\end{eqnarray}
Note that Eq.~(\ref{eq:singular-explicit}) is valid for both top quark decay $t
\rightarrow H^+ + b + X$ and charged Higgs decay $H^- \rightarrow \bar{t} + b
+ X$, which is a merit of the proper choice of the slicing parameter $\rho$
defined in Eq.~(\ref{eq:rho}).

\section{Numerical result}\label{sec:num-res}

In the numeric calculations, we use 3-loop running of $\alpha_s$ with
$\alpha_s^{(N_l = 5)} (m_Z) = 0.$1181 \cite{PDG2020}. The pole masses of the top
quark and bottom quark are set to 172.5 GeV and 4.78 GeV respectively.
For the Yukawa coupling, the $\overline{\rm MS}$ masses of top and bottom quarks
run at three-loop and match at two-loop near the flavor threshold
\cite{hep-ph/0004189}. The vacuum expectation value takes $v = 246.22$ GeV.
We set the renormalization scale to $\mu_R = \frac{1}{2} m_H$ for charged Higgs boson
decay $H^- \rightarrow \bar{t} + b + X$ and to $\mu_R = m_t$ for top quark decay $t
\rightarrow H^+ + b + X$ unless specified, as will be justified later.

We use the \textsc{Vegas} \cite{Lepage:1977sw} algorithm implemented in 
\textsc{Cuba} library \cite{hep-ph/0404043} to perform the numerical integration. 
The polylogarithm and harmonic polylogarithm functions appeared in the hard function
are calculated by \textsc{handyG} library \cite{1909.01656}.
The scalar integrals in the NLO corrections of the three-body decay are
numerically calculated with \textsc{QCDLoop} \cite{0712.1851}.
The LO matrix elements of the four-body decay are calculated with \textsc{HELAS} library
{\cite{Murayama:1992gi}}.

\subsection{Charged Higgs boson decay}\label{subsec:chagedHiggsDecay}
In this subsection, a charged Higgs boson with mass greater than the top
quark is considered. In the first part, the dependence of our predictions
on the cut-off of the phase space slicing variable and on the renormalization scale
is demonstrated. By comparison with the analytical
results available at NLO, we justify the consistency of our results 
for small cut-offs. We then show our NNLO predictions with various charged Higgs
boson masses in the second part.

\subsubsection{Validation of the calculations}

\begin{figure}[h]
  \subfloat[$m_H=300$ GeV]{\includegraphics[width=0.45\textwidth]{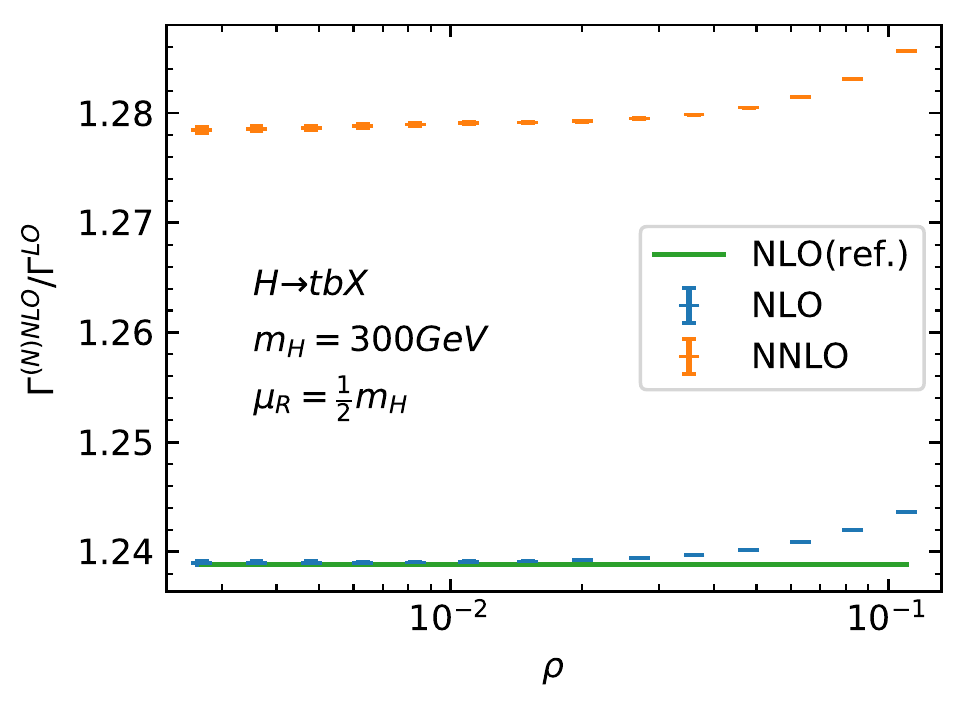}} \qquad\qquad
  \subfloat[$m_H=1500$ GeV]{\includegraphics[width=0.45\textwidth]{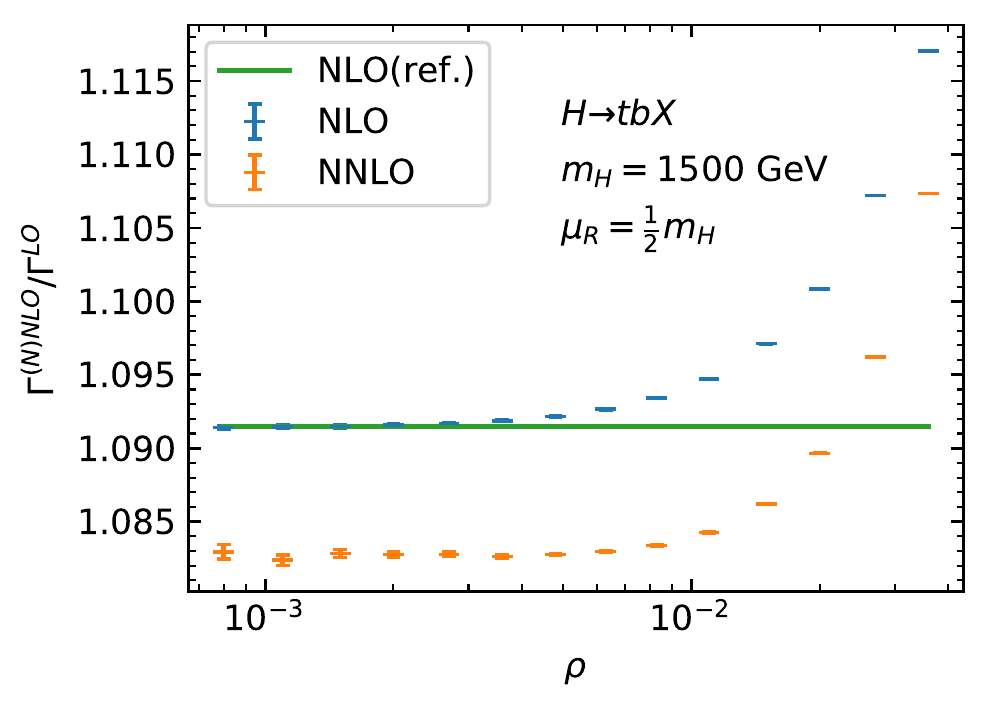}}
  \caption{\label{fig:higgs-rho}Dependence of the partial decay width on the
  phase space slicing parameter cut-off $\rho_{\rm cut}$ for a 300 GeV (left) or a 1500 GeV (right) charged Higgs boson
  decaying into a top quark and jets. The $\rho$ parameter is defined in
  Eq.~(\ref{eq:rho}). The green horizontal lines represent the analytical predictions
  of the NLO partial decay width. The blue (orange) scatter points with error bars represent
  our NLO (NNLO) predictions. The renormalization scale is set to the half of
  the charged Higgs boson mass. All predictions are normalized to the LO partial
  width $\Gamma_{H \rightarrow t b}^{\tmop{LO}}$ with m$_H$ = 300 GeV (left)
  or 1500 GeV (right)}
\end{figure}

In Fig.~\ref{fig:higgs-rho} we plot the dependence of the NLO and the NNLO
partial decay width on the cut-off of the phase space slicing variable
$\rho$ for a 300 GeV (left) or a 1500 GeV (right) charged Higgs boson decaying
into a top quark and jets. All predictions are normalized to the
LO partial width for simplicity. The blue (orange) scatter points with error bars
represent our NLO (NNLO) calculations with Monte-Carlo statistical uncertainties,
and the green horizontal lines represent the analytical NLO predictions 
shown in Eq.~(\ref{eq:NLO-higgs}).

In both panels, as $\rho$ decreases, our NLO results clearly approach
the analytical NLO predictions.
The deviations of our NLO results from the genuine predictions for large 
cut-off $\rho_{\tmop{cut}}$ are due to power corrections.
For $\rho_{\tmop{cut}}$ below 0.03 (0.005) for a 300 (1500) GeV
charged Higgs boson, however, these differences are within the Monte-Carlo statistical 
errors and are less than one per mille.
The power corrections in these regions can therefore be safely neglected.
In spite of the consistency between our numerical and the analytical results,
an increase in the Monte-Carlo uncertainties manifests for very small cut-offs,
which is especially marked for $m_H = 1500 \tmop{GeV}$ at NNLO.
Our NNLO results tend to be stable when $\rho_{\tmop{cut}}$ is small.
The Monte-Carlo uncertainties instead grows by multiple times, but still
keep small in absolute values.
Furthermore, a clear distinction between the soft or collinear region and the fixed order 
region is needed for definite predictions.
Empirically, we take $\rho_{\tmop{cut}} = 3 \tmop{GeV} / m_H$ in
the following analyses.

%%%%%%%%%%%%%%%%%%%%%%%%%%%%%%%%%%%%%%%%%%%%%%%%%%%%%%%%%%%%%%%%%%%%%%%%%%%%%

\begin{figure}[h]
\centering
  \raisebox{0.0\height}{\includegraphics[width=0.55\textwidth]{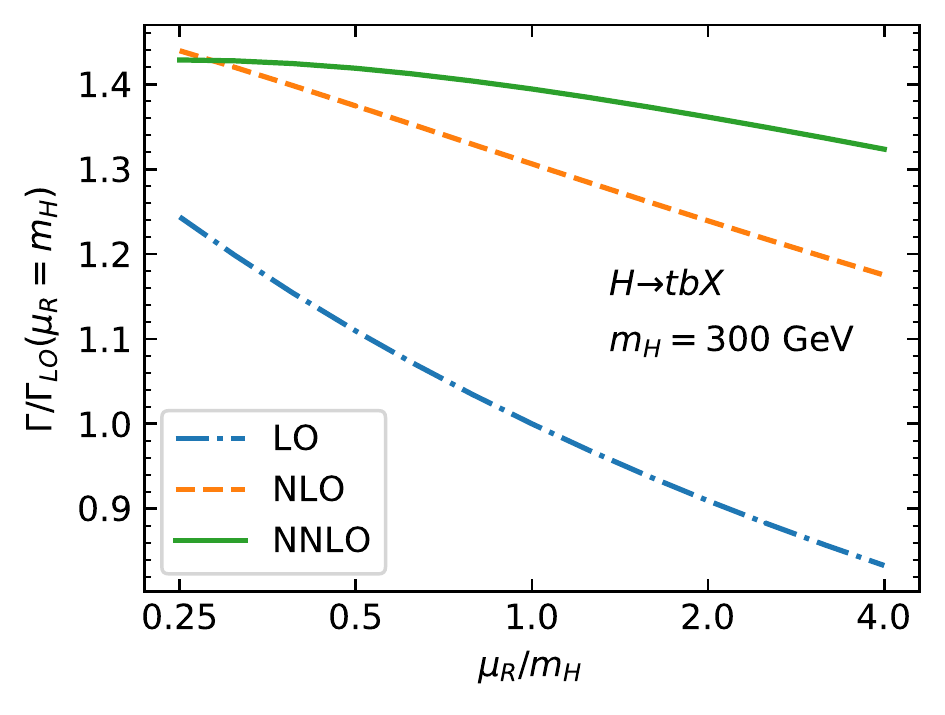}}
  \caption{\label{fig:higgs-mu}Renormalization scale dependence of the 
  LO (dot-dashed blue line), NLO (dashed orange line) and NNLO (green line) partial
  decay width of a 300 GeV charged Higgs boson decaying into
  a top quark and jets. All results are normalized to the LO partial width
  at central scale $\mu_R = m_H$. Monte-Carlo uncertainties are rather small
  and are not shown here.}
\end{figure}

The dependence on renormalization scale for the partial decay width of a 300 GeV
charged Higgs boson is shown in Fig.~\ref{fig:higgs-mu}.
In this figure, predictions at LO, NLO and NNLO are plot in dot-dashed blue,
dashed orange and green lines, respectively. 
All results are normalized to the LO partial width at central scale $\mu_R = m_H$.
As we can see, the LO
partial decay width varies by about $-17\% \sim 24$\% as the renormalization scale
changes by a factor of 4. The dependence at LO is completely due to the running of the
$\overline{\rm MS}$ Yukawa couplings. The NLO partial decay width has an
uncertainty of about $-13\% \sim 13$\%. 
The renormalization scale uncertainty is further
suppressed to about $-7\% \sim 3$\% of the LO width
with the inclusion of the NNLO QCD corrections.
Besides, from Fig.~\ref{fig:higgs-mu} it indicates
that taking 
 $\mu_R = m_H/2$ or $\mu_R = m_H/4$ leads to better convergences of the perturbative series. 
The optimal renormalization scale actually depends on the mass of
the charged Higgs boson. As is mentioned above,
a general setting of $\mu_R = m_H/2$ is 
used for $200{\rm GeV} \le m_H \le 3000{\rm GeV}$
in our work, unless otherwise specified.

%%%%%%%%%%%%%%%%%%%%%%%%%%%%%%%%%%%%%%%%%%%%%%%%%%%%%%%%%%%%%%%%%%%%%%%%%%%%%
\subsubsection{NNLO partial width for different $m_H$}

\begin{figure}[h]
\centering
  \raisebox{0.0\height}{\includegraphics[width=0.65\textwidth]{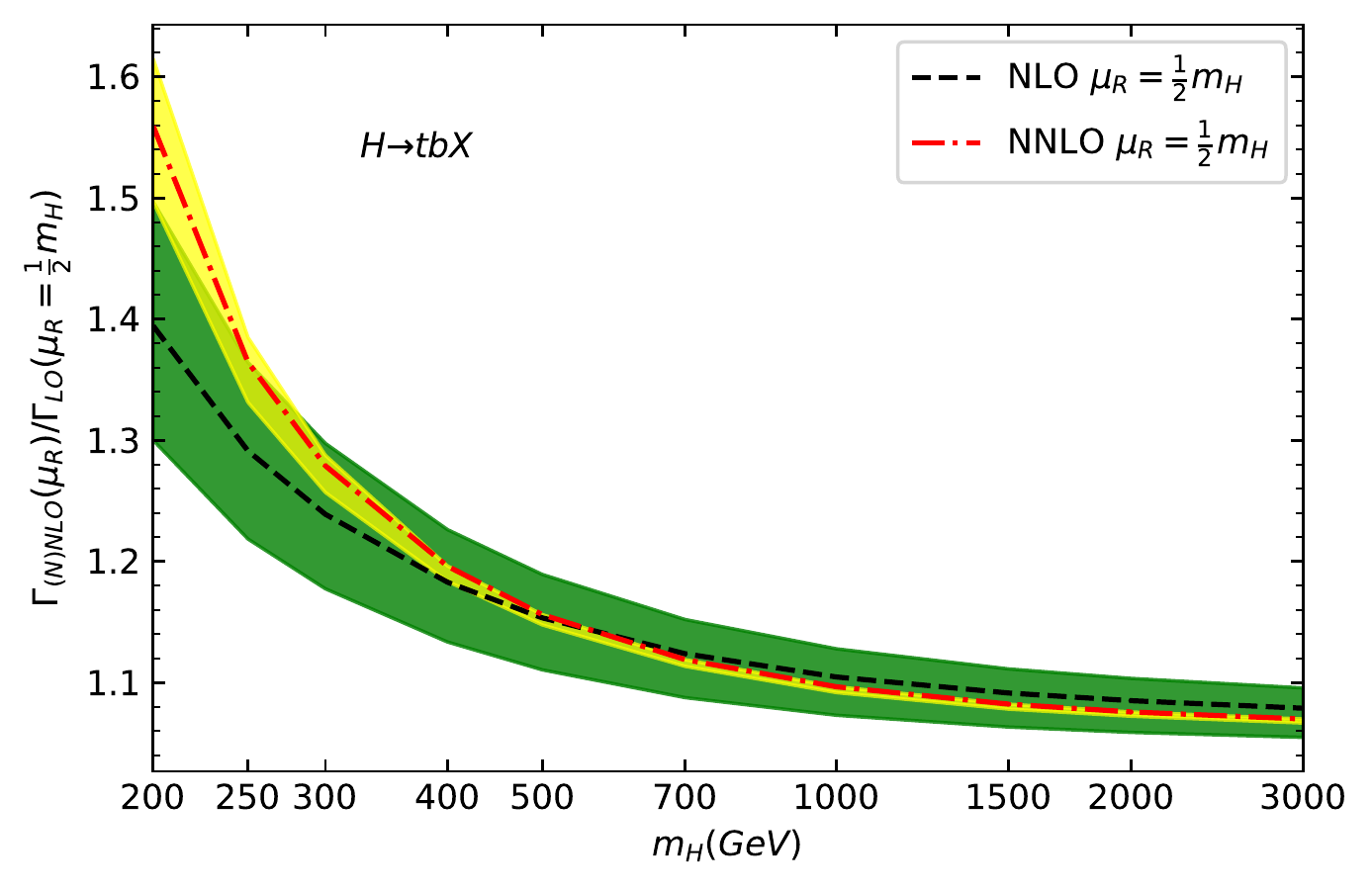}}
  \caption{\label{fig:higgs-mh}The NLO (dashed black line) and NNLO (dot-dashed red
  line) partial decay width of charged Higgs boson
  decaying into a top quark and jets for different charged Higgs boson masses.
   The green (yellow) band is bounded by the maximum and minimum of NLO
  (NNLO) partial width at three renormalization scales
  $\mu_R = m_H, \frac{1}{2} m_H, \frac{1}{4} m_H$.
  All the results are normalized to the LO partial width at $\mu_R = \frac{1}{2}m_H$.}
\end{figure}

In Fig.~\ref{fig:higgs-mh}, we demonstrate the dependence of the partial decay width 
on the mass of the charged Higgs boson for its decay to a top quark
and jets. 
In this figure, the NLO (NNLO) partial decay width is plotted in dashed black (dot-dashed
red) line.
The green band
  is bounded by the NLO partial width at three different renormalization scales,
  more clearly, it is between the minimum and the maximum of 
    $ \left\{ \Gamma^{\tmop{NLO}} \left( \mu_R =
         \mu \right)  / \Gamma^{\tmop{LO}} \left( \mu_R =
         m_H/2 \right) \right\}$, for  $\mu = m_H/4$, $m_H/2$  or $m_H $
  , and the yellow band is the NNLO counterpart. These two
  bands give estimations of the residual perturbative uncertainties.
  Though the possibility exists that 
  the ratios 
    locate outside the green (yellow) band for certain $\mu \in
  \left[ \frac{1}{4} m_H, m_H \right]$.
All these results are normalized to $\Gamma^{\tmop{LO}} (\mu_R = m_H/2)$.
Compared with the NLO results, the uncertainties from renormalization scale are 
significantly reduced in entire ranges of the charged Higgs boson masses considered
once the NNLO corrections are included.
For a charged Higgs boson with moderate mass of about 400 GeV, this reduction can be as
large as 80\%.
Moreover, the NNLO corrections are most sizable for charged Higgs boson with low masses.
For a charged Higgs boson with mass of 200 GeV, the corrections reach up to 10\%, and
the corrections decrease to about 1\% for a mass of 3000 GeV.

We also summarize 
ratios of the NNLO (NLO) to the LO predictions
using the same renormalization scale in both the numerator and the denominator,
for a variety of the charged Higgs boson masses in Table~\ref{tab:higgs-mh}.
As we already commented at the end of Sec.\ref{sec:eff-oper},
such ratios are independent of the detailed structure of the Yukawa couplings,
and may be applied to different types of two-Higgs-doublet models.
The renormalization scale is chosen as either the mass of the charged Higgs
or the half of that.
At $\mu_R = m_H$, the NNLO corrections decrease monotonically with the mass of the charged Higgs
boson, from 22\% for a mass of 200 GeV to about 1\% for 3000 GeV.
The change of the scale from $m_H$ to $m_H/2$ has little impact on this trend, and
leads to smaller corrections in general.

\begin{table}[h]
\renewcommand\arraystretch{1.0}
\centering
\begin{tabular}{|c||c|c|c|c|c|c|c|c|c|c|}
\hline
$m_H$/GeV              &200&250&300&400&500&700&1000&1500&2000&3000\\\hline\hline
NLO($\frac{1}{2}m_H$)  &1.394&1.291&1.239&1.183&1.153&1.124&1.104&1.091&1.085&1.079\\\hline
NNLO($\frac{1}{2}m_H$) &1.559&1.365&1.279&1.196&1.156&1.119&1.096&1.082&1.076&1.070\\ \hline\hline
NLO($m_H$)             &1.451&1.355&1.306&1.253&1.224&1.194&1.173&1.158&1.150&1.142\\\hline
NNLO($m_H$)            &1.670&1.481&1.394&1.308&1.265&1.222&1.195&1.175&1.166&1.156\\ \hline
\end{tabular}
\caption{Ratios $\Gamma^{\rm (N)NLO}(\mu_R)/\Gamma^{\rm LO}(\mu_R)$ for $\mu_R=m_H/2 \text{ or } m_H$. Monte-Carlo statistical uncertainties at NNLO are small and not shown.}
\label{tab:higgs-mh}
\end{table}

Additionally, throughout our calculation, a five-flavor strong
coupling is used and the Yukawa coupling is renormalized
in $\overline{\rm MS}$ scheme. Though the Yukawa coupling is canceled out
in the ratio $\Gamma^{\rm NNLO(NLO)}(\mu_R)/\Gamma^{\rm LO}(\mu_R)$, the numeric value of this ratio
will be quite different
if an on-shell Yukawa coupling is taken.
The relation between Yukawa couplings under these two schemes can be found in
Appendix \ref{app:FO}.

%%%%%%%%%%%%%%%%%%%%%%%%%%%%%%%%%%%%%%%%%%%%%%%%%%%%%%%%%%%%%%%%%%%%%%%%%%%%%
%%%%%%%%%%%%%%%%%%%%%%%%%%%%%%%%%%%%%%%%%%%%%%%%%%%%%%%%%%%%%%%%%%%%%%%%%%%%%
\subsection{Top quark decay}
In this subsection, a charged Higgs boson with mass smaller than the top quark is considered. 
Converse to the case in subsection~\ref{subsec:chagedHiggsDecay}, the light charged Higgs
boson now turns out to be the decay product of the top quark. Analogous discussions to 
that subsection are performed.

\subsubsection{Validation of the calculations}
\begin{figure}[h]
  \subfloat[$m_H=100$ GeV]{\includegraphics[width=0.45\textwidth]{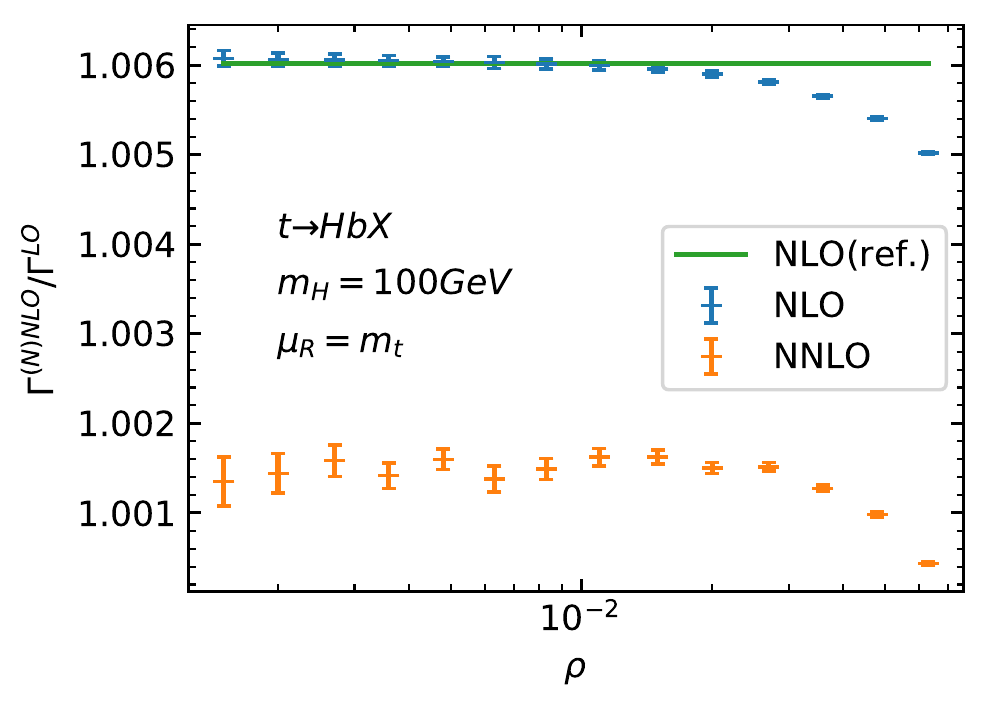}} \qquad\qquad
  \subfloat[$m_H=140$ GeV]{\includegraphics[width=0.45\textwidth]{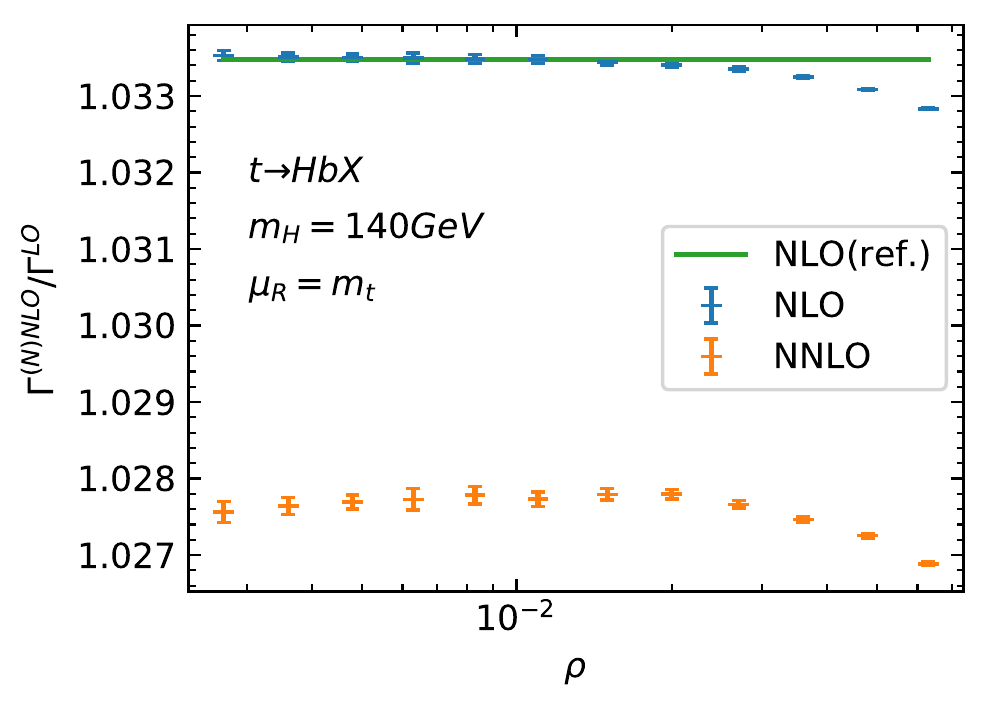}}

  \caption{\label{fig:top-rho}
  Dependence of the partial decay width on the
  phase space slicing parameter cut-off $\rho_{\rm cut}$ for the top quark decaying into
  a 100 GeV (left) or a 140 GeV (right) charged Higgs boson and jets. The $\rho$ parameter is defined in
  Eq.~(\ref{eq:rho}). The green horizontal lines represent the analytical predictions
  of the NLO partial decay width. The blue (orange) scatter points with error bars represent
  our NLO (NNLO) predictions. The renormalization scale is set to the pole
  mass of the top quark. All predictions are normalized to the LO partial
  width $\Gamma_{t \rightarrow H b}^{\tmop{LO}}$ with $m_H$ = 100 GeV (left) or 140 GeV (right).}
\end{figure}

Given a light charged Higgs boson, the top quark can decay into the charged Higgs boson
along with jets.
In Fig.~\ref{fig:top-rho}, we show the dependence of NLO (blue) and NNLO (orange) partial decay width on the cut-off
of the phase space slicing parameter, $\rho_{\rm cut}$, for a 100 GeV (left) or a 140 GeV (right)
charged Higgs boson. The green horizontal lines represent the analytical NLO predictions given in Eq.~(\ref{eq:NLO-top}).
All predictions are normalized to the LO partial width.

As with the case in subsection~\ref{subsec:chagedHiggsDecay}, for small enough cut-offs,
the deviations of our NLO predictions from the analytical NLO calculations are negligible.
The empiric choice of $\rho_{\tmop{cut}} = 3 \tmop{GeV} / m_H$ is, however, no longer valid in
current scenarios.
For a charged Higgs mass of 100 GeV, the empiric value is 0.03 which is away from the stable
region. Considering the range of the charged Higgs boson mass studied is limited, 
in the following analyses we choose a constant value of $\rho_{\rm cut} = 0.01$
which is small enough to give a stable result.

Under this convention, we can find that the difference between the genuine results
and our predictions at NLO is within a per mille.
At NNLO, instead, the relative fluctuations in the stable region, and the Monte-Carlo uncertainties seem to be large especially 
for a charged Higgs boson with mass of 100 GeV,
which is due to the fact that higher-order corrections are small, while the Monte-Carlo
uncertainties are only relevant to the size of generating samples.
Considering the absolute deviations, we can safely neglect the Monte-Carlo errors as well as the
power corrections.

\begin{figure}[h]
\centering
  \raisebox{0.0\height}{\includegraphics[width=0.55\textwidth]{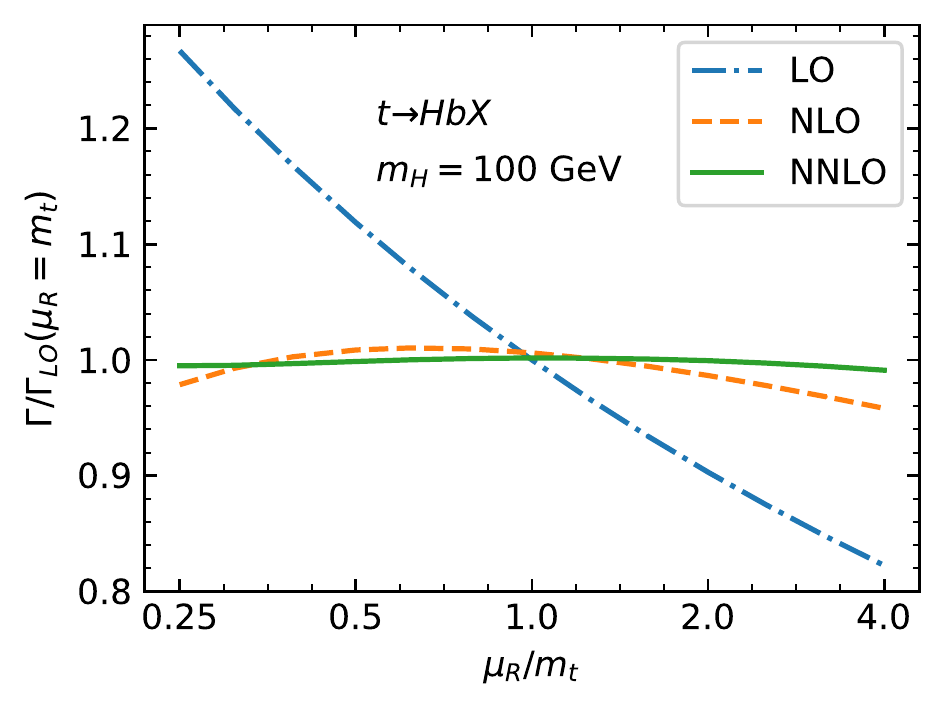}}
  \caption{\label{fig:top-mu}Renormalization scale dependence of the LO (dot-dashed blue line), NLO
(dashed orange line) and NNLO (green line) partial decay width of a  top quark decaying into a 100 GeV charged Higgs boson
 and jets. All the results are normalized to the LO partial width
  at central scale $\mu_R = m_t$. The Monte-Carlo uncertainties are very small
  and are not shown here.}
\end{figure}

%%%%%%%%%%%%%%%%%%%%%%%%%%%%%%%%%%%%%%%%%%%%%%%%%%%%%%%%%%%%%%%%%%%%%%%%%%%%%
In Fig.~\ref{fig:top-mu}, we show the renormalization scale dependence of the
partial width of the top quark decaying into a 100 GeV charged Higgs
boson. Results at LO, NLO and NNLO are plotted in dot-dashed blue, dashed orange
and green lines, respectively. All the results are normalized to the LO partial width
at a central scale $\mu_R = m_t$. 
As can be seen, the introduction of higher-order corrections significantly reduces
the renormalization scale dependence already at NLO.
The supplement of NNLO corrections further stabilize the predictions for $\mu_R \in
\left[ \frac{1}{4} m_t, 4m_t \right]$ to within 1\%, which indicates the importance of 
higher-order corrections in this situation. From Fig.~\ref{fig:top-mu} it indicates
the optimal renormalization scale choice is $m_t$ which shows a very good perturbative convergence.

\subsubsection{NNLO partial width for different $m_H$} \label{sec:top-mh-dep}

\begin{figure}[h]
\centering
  \raisebox{0.0\height}{\includegraphics[width=0.65\textwidth]{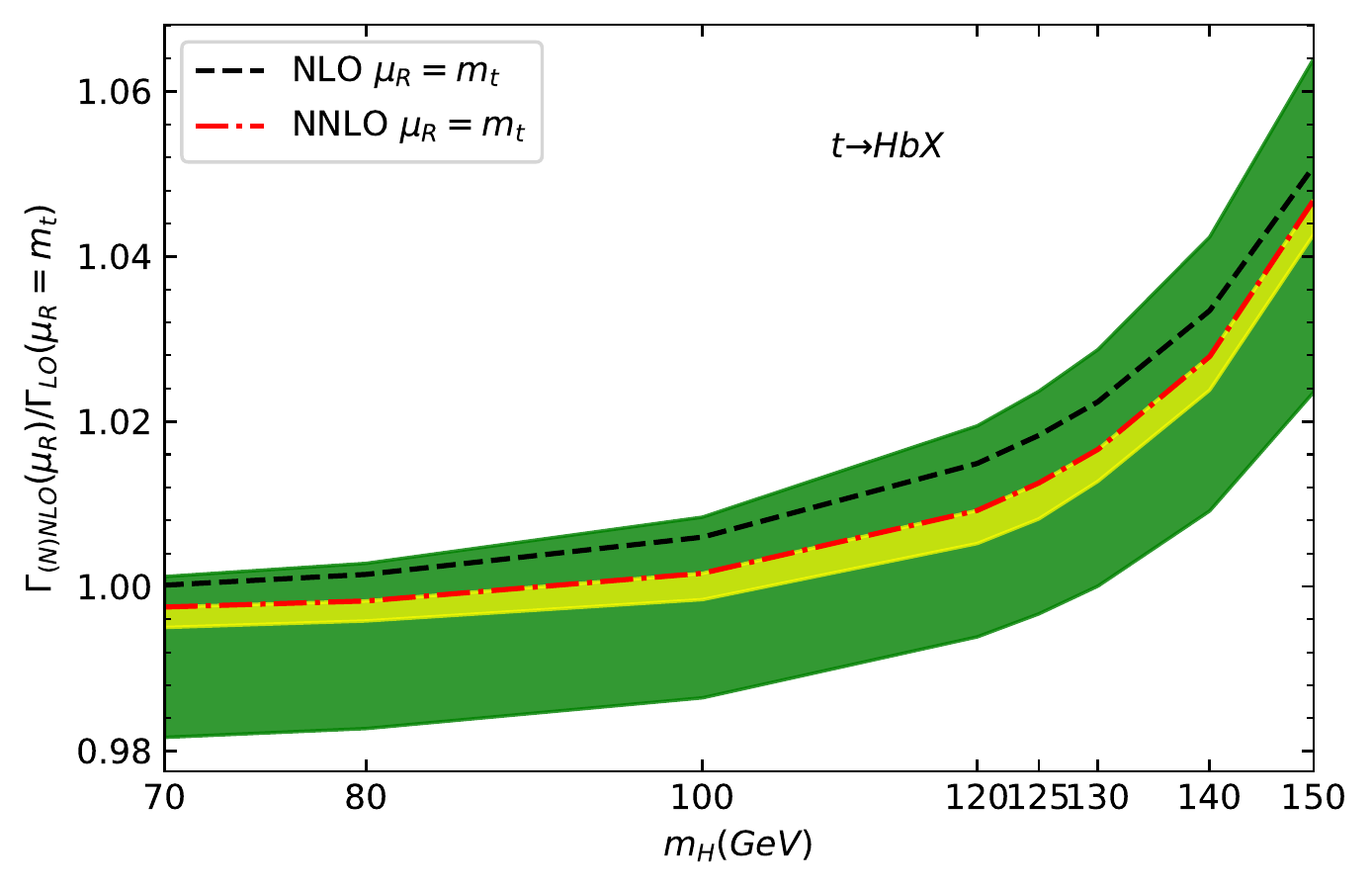}}
  \caption{\label{fig:top-mh}
  The NLO (dashed black line) and NNLO (dot-dashed red
  line) partial decay width of top quark decaying into a
  charged Higgs boson and jets for different charged Higgs boson masses.
   The green (yellow) band is bounded by the maximum and minimum of NLO
  (NNLO) partial width at three renormalization scales
  $\mu_R = \frac{1}{2} m_t, m_t, 2 m_t$.
  All the results are normalized to the LO partial width at $\mu_R = m_t$.}
\end{figure}

The variation of the partial decay width with the charged Higgs boson mass, as 
well as bands of scale variations, are shown in
Fig.~\ref{fig:top-mh}, with all results normalized to $\Gamma^{\rm LO}({\mu_R=m_t})$. 
Our NLO (NNLO) results are plotted in dashed black (dot-dashed red) line.
On contrary to the tens of percent level corrections to the LO results for a heavy charged Higgs boson,
QCD corrections now are at most a few percents.
Nevertheless, as we have shown in Fig.~\ref{fig:top-mu}, the supplement of these
higher-order corrections significantly reduces the scale uncertainties.
In most regions, the incorporation of NNLO corrections can further reduce the NLO scale uncertainties
by at least 70\%.

\begin{table}[h]
\renewcommand\arraystretch{1.0}
\centering
\begin{tabular}{|c||c|c|c|c|c|c|c|c|}
\hline    
$m_H$/GeV              &70&80&100&120&125&130&140&150\\\hline\hline
NLO($\frac{1}{2}m_t$)  &0.895&0.896&0.901&0.911&0.915&0.920&0.932&0.951\\\hline
NNLO($\frac{1}{2}m_t$) &0.890&0.890&0.893&0.899&0.901&0.905&0.915&0.932\\\hline\hline
NLO($m_t$)             &1.000&1.001&1.006&1.015&1.018&1.022&1.033&1.051\\\hline
NNLO($m_t$)            &0.997&0.998&1.002&1.009&1.013&1.017&1.028&1.047\\\hline
\end{tabular}
\caption{Ratios $\Gamma^{\rm (N)NLO}(\mu_R)/\Gamma^{\rm LO}(\mu_R)$ for $\mu_R= m_t/2\text{ or }m_t$. Monte-Carlo statistical uncertainties at NNLO are small and not shown.}
\label{tab:top-mh}
\end{table}

Analogously, the ratios of the NNLO or the NLO to the LO predictions at the renormalization
scale of $m_t$ or $m_t/2$ are summarized in Table \ref{tab:top-mh}. 
Unlike the case of a heavy charged Higgs boson, the NNLO corrections are rather small.
At $\mu_R=m_t$, NNLO corrections over the entire range of the charged Higgs boson mass are at per mille
level.
On the other hand, difference due to the change of scale can be sizable, which mainly originates from the large
scale uncertainties of the denominator $\Gamma^{\rm LO}(\mu_R)$.
This can be deduced in combination with Fig.~\ref{fig:top-mu}
in which the running coupling of the denominator is fixed at scale $m_t$.
Also, it is worthy noting that, all scale variances at LO are from the running of the Yukawa couplings. 

\subsection{Phenomenological implications}
In this subsection, we show some phenomenological implications of our results for typical two-Higgs-doublet models. The Higgs sector of the two-Higgs-doublet models consists of two $SU(2)_L$ scalar doublets $\Phi_i(i=1,2)$ with hyper-charge $Y=1/2$~\cite{1106.0034},

\begin{eqnarray}
	 \Phi_i= \left ( \begin{array}{cc}
		\phi_i^{+}\\
		(v_{i}+\phi_i^0+iG_i^0)/\sqrt{2}
	\end{array}
	\right )\;,
\end{eqnarray}
where the $\phi_i^+,\phi_i^0$ and $G_i^0(i=1,2)$ are the parametrized  component fields. $v_i(i=1,2)$ are the vacuum expectation values of the doublets after the electroweak symmetry breaking, satisfying $\sqrt{v_1^2+v_2^2}=v=246$GeV. 
  There are eight degrees of freedom with the two complex scalar $SU(2)_L$ doublets. Three of those give mass to the $W^{+},W^{-}$ and $Z^0$ gauge bosons from the Higgs mechanism, the remaining five are physical scalar (‘Higgs’) fields. There are two charged scalars, two neutral scalars, and one pseudoscalar.

The Lagrangian for the Higgs sector of the two-Higgs-doublet models is written as following:

\begin{equation}
	\mathcal{L}=\sum_{i}|D_{\mu}\Phi_{i}|^2-V(\Phi_{1},\Phi_{2})+\mathcal{L}_{Yuk},
\end{equation}
where $D_{\mu}$ is the covariant derivative and $V(\Phi_1,\Phi_2)$ is the scalar potential. The Yukawa sector of the two Higgs doublets is given by 
  \begin{equation}
  	-\mathcal{L}_{Yuk}=Y_{u}\bar Q_{L}\tilde{\Phi}_uu_R+Y_{d}\bar Q_{L}\Phi_dd_R+Y_{e}\bar L_{L}\Phi_ee_R+h.c.,
  \end{equation}
where $\tilde{\Phi}=i\sigma_2\Phi^*$; ${Q}_L$ and ${L}_L$ are the quark and the lepton left-hand doublet; $u_{R},d_{R},e_{R}$ are the right-hand singlet; $Y_{u,d,e}$ are Yukawa coupling constants; $\Phi_{u,d,e}$ are either $\Phi_1$ or $\Phi_2$. To avoid tree-level flavor-changing-neutral-currents, a discrete $\mathbb{Z}_2$ symmetry is imposed~\cite{Han:2020lta}. There are four possible choices for the charge
assignment of the fermions under $\mathbb{Z}_2$, corresponding to type-I, type-II, type-X and type-Y respectively. We summarize the charge assignment of the four types of 2HDM in Table.\ref{types}, along with the non-zero Yukawa couplings for each $\Phi$.
\begin{table}[h]
	\centering
	\begin{tabular}{ccccccccc}
		\hline\hline
		Types   & $\Phi_1$ & $\Phi_2$ & $u_R$ & $d_R$ & $l_R$ & $Q_L,L_L$ & $\Phi_1$ & $\Phi_2$ \\ \hline
		Type-I  & $+$        & $-$        & $-$     & $-$     & $-$     & $+$         &          & $u,d,l$  \\
		Type-II & $+$        & $-$        & $-$     & $+$     & $+$     & $+$         & $d,l$    & $u$      \\
		Type-X  & $+$        & $-$        & $-$     & $-$     & $+$     & $+$         & $l$      & $u,d$    \\
		Type-Y  & $+$        & $-$        & $-$     & $+$     & $-$     & $+$         & $d$      & $u,l$    \\ \hline\hline
	\end{tabular}
\caption{
	Four types of assignments for the $\mathbb{Z}_2$ charges of the $\Phi_{1,2}$ and SM fermions. The last two columns indicate the non-zero Yukawa coupling of each scalar doublets $\Phi_{1,2}$.}
\label{types}
\end{table}

   In the following we focus on the type-II and type-X of the two-Higgs-doublet models. We set the model parameters to be $\tan\beta=20$, $\sin(\beta-\alpha)=0.995$ and $\cos(\beta-\alpha)>0$ according to the benchmark point in Ref.~\cite{Aiko:2021can}. In addition, we set the mass of charged Higgs boson($H^{+}$) and the neutral Higgs boson ($A,H$) to be the same, varying from $200$ GeV to $1500$ GeV. Moreover, the renormalization scale is chosen as the mass of the Higgs boson. We study the branching ratio of the charged Higgs boson decaying into the top quark and the anti-bottom quark, which can be expressed as

\begin{equation}
	BR(H^{+}\to t\overline{b})=\frac{\Gamma_{t\bar{b}}}{\Gamma_{t\bar{b}}+\Gamma_{rest}},
\end{equation}
where $\Gamma_{t\bar{b}}$ can be the LO, NLO and NNLO decay width of the channel  $H^{+}\to t\overline{b}$ presented earlier. 
$\Gamma_{rest}$ represents partial width from all other decay channels which we calculated with  2HDMC-1.8.0~\cite{Eriksson:2009ws},
where LO QCD corrections are included if the charged Higgs boson decays to two massless quarks, 
while all other channels only include the Born level contributions.

\begin{figure}[h]
\centering
\raisebox{0.0\height}{\includegraphics[width=8.cm,height=4.8cm]{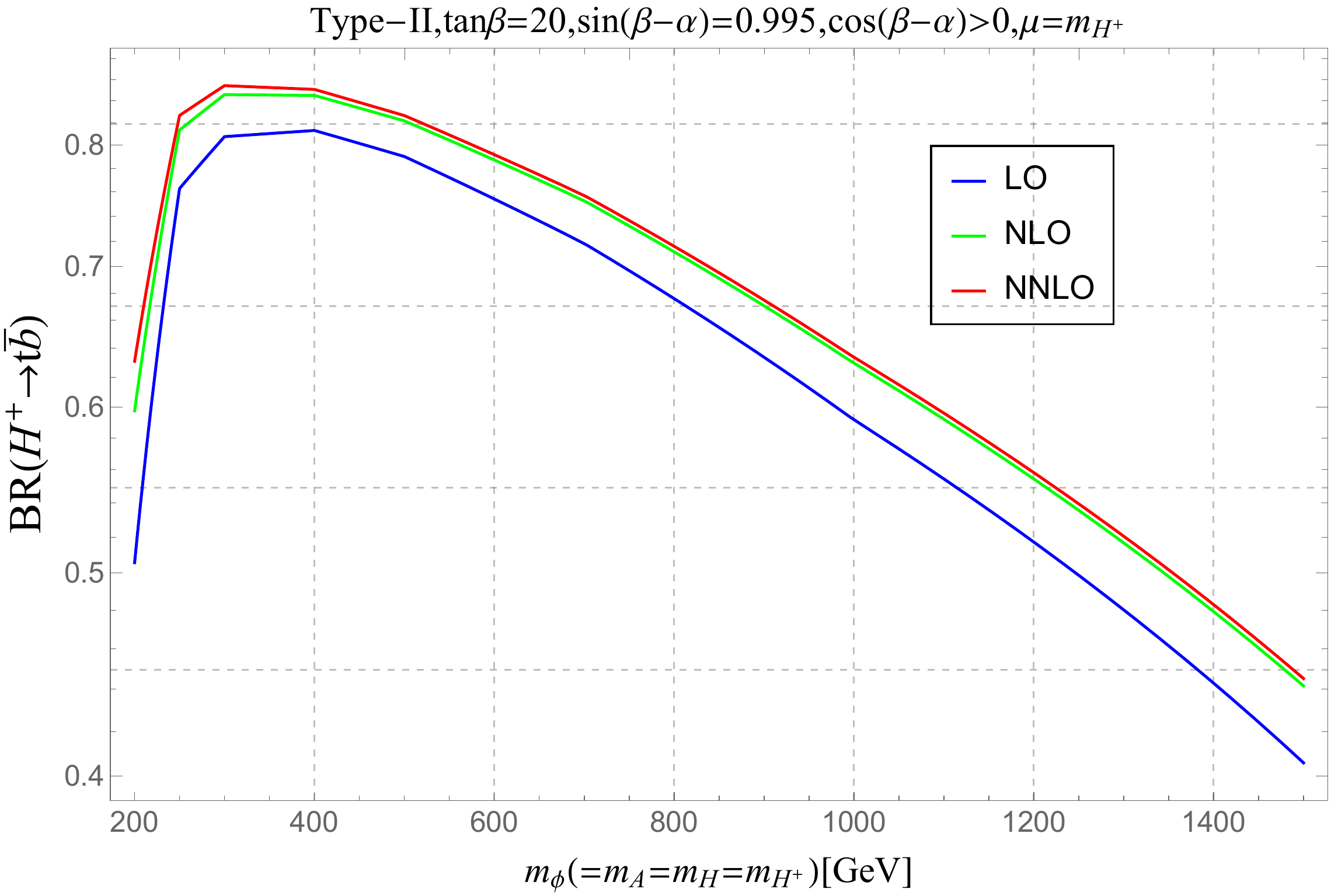}}
\caption{
Branching ratio of the charged Higgs boson decay into $t\bar b$ in the 2HDM of Type-II , using the LO(Blue), NLO(Green) and NNLO(Red) partial width with $\tan\beta=20$, $\sin(\beta-\alpha)=0.995$ and $\cos(\beta-\alpha)>0$.}
\label{brII}
\end{figure}

In Fig.~\ref{brII}, we show the branching ratio as a function of the
Higgs boson mass at different orders of QCD for the type-II model. One
can find that the branching ratio firstly increases then
decreases. The increase of the branching ratio is because that the
phase space of the channel $H^{+}\to t\bar{b}$ is increasing, the
Yukawa coupling of the charged Higgs boson with the top and bottom
quark is large. The branching ratio decreases at large values of the
charged Higgs boson mass due to the contribution of the charged Higgs
boson decaying into the W boson and the SM Higgs boson
($H^{+}\to W^{+}h$) to the total width. The branching ratio is $60\%$
and $63\%$ at the NLO and NNLO respectively, for a charged Higgs boson
mass of $200$ GeV.
The observed (expected) $95\%$ confidence level (CL) upper limits of
$\sigma(pp\to tbH^{+})\times \mathcal{B}(H^{+}\to tb)$ range from
3.6(2.6)pb at $m_{H^{+}}=200$GeV to 0.036(0.019)pb at
$m_{H^{+}}=2$TeV~\cite{2102.10076}. It shows that the observed limits
are improved by $5\%$ to $70\%$ depending on the mass of the charged
higgs boson. And the relative accuracy between
the NLO and NNLO predictions given by our results is $5\%$ at $m_{H^{+}}=200$GeV.
The NLO and NNLO predictions are both about $45\%$
when the mass is $1500$ GeV. Overall, one can find that the NNLO
corrections are most significant when the charged Higgs mass is
between $200$ GeV and $500$ GeV, and fade away while the mass
increases. The NNLO corrections are much smaller than the NLO
corrections indicating a good convergence of the perturbative
calculations. 
\begin{figure}[h]
\centering
\raisebox{0.0\height}{\includegraphics[width=8.cm,height=4.8cm]{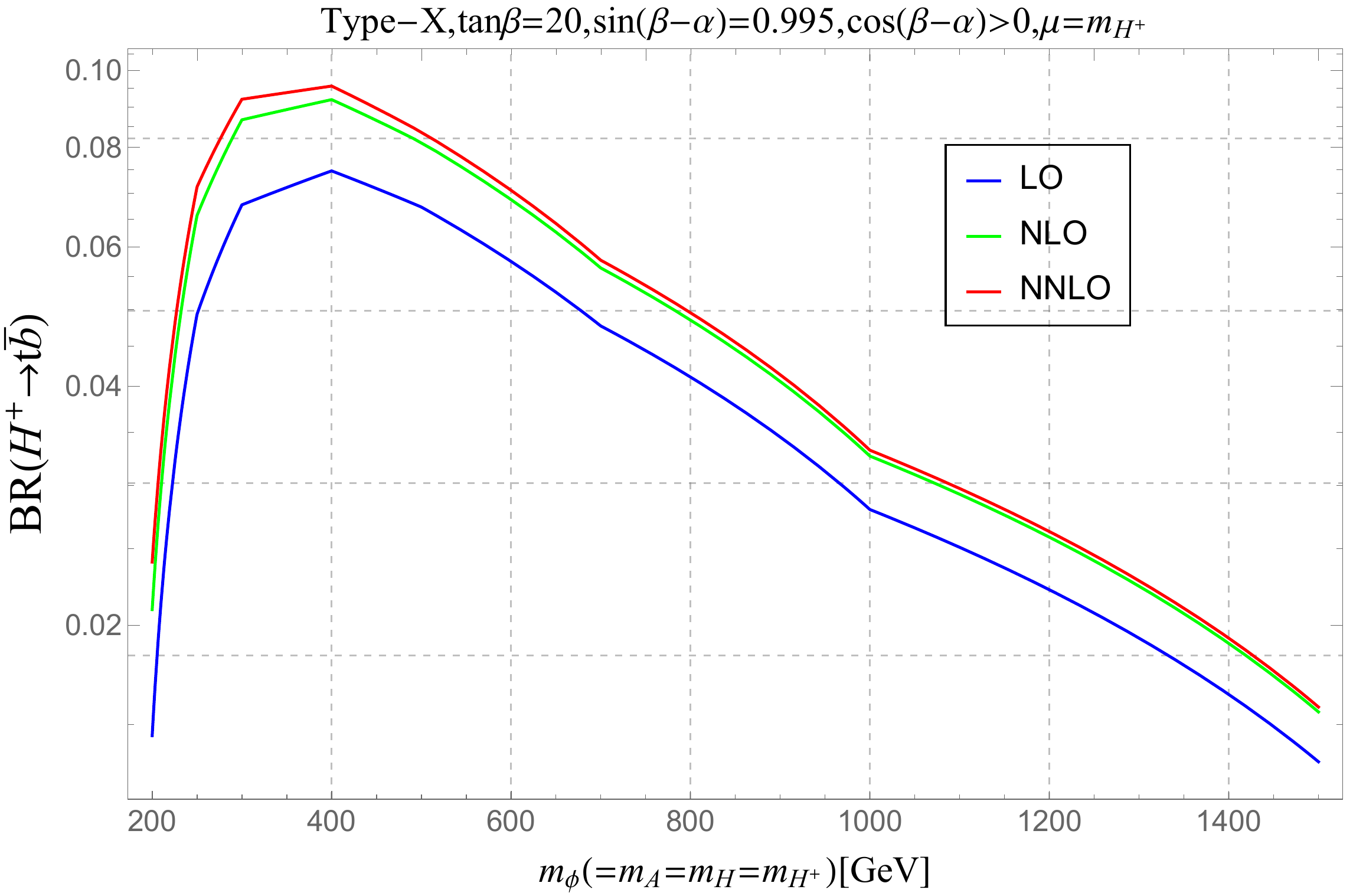}}
\caption{
Branching ratio of the charged Higgs boson decay into $t\bar b$ in the 2HDM of Type-X, using the LO(Blue), NLO(Green) and NNLO(Red) partial width with $\tan\beta=20$, $\sin(\beta-\alpha)=0.995$ and $\cos(\beta-\alpha)>0$.}
\label{brX}
\end{figure}

     We present results of the branching ratio for the type-X model in Fig.~\ref{brX}. We can see that the branching ratio shows a similar dependence on the charged Higgs boson mass but is much smaller. It is because the Yukawa coupling constant is different between for the two models. The branching ratio is $2.1\%$ and $2.4\%$ at the NLO and NNLO respectively, for a charged Higgs boson mass of $200$ GeV. The NLO and NNLO predictions are both about $1.6\%$ when the mass is $1500$ GeV.

  The $95\%$ confidence level upper limits of
    $\sigma(pp\to tbH^{+})\times \mathcal{B}(H^{+}\to tb)$ is observed by
    the ATLAS collaboration, ranging from 3.6~pb at $m_{H^{+}}=200$ GeV to
    0.036~pb at $m_{H^{+}}=2$ TeV~\cite{2102.10076}. Compared to the
    previous ATLAS search, the limits is improved by $5\%$ to $70\%$
    depending on the mass of the charged higgs boson.

\section{Summary}\label{sec:summary}
In this work, we present the calculations of NNLO partial width of the charged Higgs boson decay
$H^- \rightarrow \bar{t} + b + X$ and the top
quark decay $t \rightarrow H^+ + b + X$ using a phase space slicing
method with the jet mass as { a slicing parameter}.
The ratios of the NNLO partial decay width to the LO prediction for a variety of
charged Higgs boson mass are given. They are independent of the
detailed structure of the Yukawa couplings, thus for a specific new
physics model, one can easily get the corresponding NNLO partial decay
width {by rescaling with the LO prediction (the Yukawa coupling should be renormalized in
$\overline{\rm MS}$ scheme).}
The renormalization scale dependence is studied in details for $m_H=$ 300GeV
(charged Higgs boson decay) and $m_H=$ 100GeV (top quark decay)
respectively.
For both cases, the renormalization scale dependence is {significantly}
reduced { by the NNLO corrections }.
The renormalization scale dependence of the NNLO partial decay width of the charged
Higgs boson is very weak for
$\mu_R=m_H/4 \sim m_H/2$, while
for the top quark decay $t \rightarrow H^+ + b + X$,
the renormalization scale { uncertainty} is within 1\%.

For the heavy charged Higgs boson decay, the NNLO corrections with $\mu_R=m_H / 2$ are
about 16\% (1\%) of the LO widths for the charged Higgs boson mass $m_H$=200~GeV (2000~GeV).
On the other hand, the NNLO corrections
for the top quark decaying into a light charged Higgs boson with $\mu_R=m_t$ are
quite small,
at about 0.3\% (0.4\%) for $m_H=$70~GeV (150~GeV).
Note the size of the QCD corrections 
can have a large renormalization scale dependence due to the running of the Yukawa couplings.
We also show some phenomenology results on the impact of our calculations to the branching ratio
of the charged Higgs boson decay in two typical 2HDM. 
The NNLO corrections increase the decay
branching ratio moderately for smaller charged Higgs boson masses.

\section*{Acknowledgments}
This work is sponsored by the National Natural
Science Foundation of China under the Grant No. 11875189 and No.11835005.
We thank the sponsorship from Yangyang Development Fund. 
We would like to thank Zelong Liu for 
proofreading the manuscript and for many valuable comments.
Zelong Liu also contributed a lot at the early stage of this work.

\appendix
\section{Ingredients of the fixed-order calculation}\label{app:FO}

The strong coupling constant $\alpha_s$ is generally renormalized in the
$\overline{\rm MS}$ scheme, and is related to the bare coupling constant
$\alpha_0$ by
\begin{eqnarray}
  \alpha_0 & = & S_{\epsilon}^{- 1} \mu^{2 \epsilon} Z_{\alpha} \alpha_s
  \nonumber\\
  & = & \alpha_s S_{\epsilon}^{- 1} \mu^{2 \epsilon} \left[ 1 + \left( -
  \frac{\beta_0}{\epsilon} \right) \frac{\alpha_s (\mu)}{4 \pi} + \left(
  \frac{\beta_0^2}{\epsilon^2} - \frac{\beta_1}{2 \epsilon} \right) \left(
  \frac{\alpha_s}{4 \pi} \right)^2 +\mathcal{O} (\alpha^3_s) \right]  \,,
\end{eqnarray}
where $\epsilon = \frac{4 - d}{2}$ is the dimensional regulator, $S_{\epsilon} =
e^{\epsilon (\ln 4 \pi - \gamma_{_E})}$, and the expansion coefficients for
the QCD beta function up to three-loop order are
\begin{eqnarray}
  \beta_0 & = & \frac{11}{3} C_A - \frac{4}{3} N_f T_F \,,\nonumber\\
  \beta_1 & = & \frac{34}{3} C_A^2 - \frac{20}{3} C_A N_f T_F - 4 C_F N_f T_F \,, \\
  \beta_2 & = & \frac{2857}{54} C_A^3 + \left( 2 C_F^2 - \frac{205}{9} C_F C_A
  - \frac{1415}{27} C_A^2 \right) T_F N_f + \left( \frac{44}{9} C_F +
  \frac{158}{27} C_A \right) T_F^2 N_f^2 \,,\nonumber
\end{eqnarray}
where $N_f$ is the number of active quark flavors, $C_A = 3, C_F = 4 / 3, T_F
= 1 / 2$ for QCD. We also use $a_s = \alpha_s (\mu) / (4 \pi)$ for simplicity.
In this work, the partial decay width is expanded in a five-flavor strong
coupling constant, which is related to the $\overline{\rm MS}$ strong
coupling constant with $N_f = 6$ by the decoupling relation Eq.~(\ref{eq:alpha-dec}).

\

The Yukawa coupling is renormalized in $\overline{\rm MS}$ scheme throughout our
calculation. The corresponding renormalization constant and anomalous
dimension can be expanded in $\alpha_s^{(N_f)}$ as \cite{hep-ph/9703284,
hep-ph/9703278, hep-ph/0411261}
\begin{eqnarray}
  Z_y = 1 & - & \frac{3 C_F}{\epsilon} \left( \frac{\alpha_s}{4 \pi} \right)
  \nonumber\\
  & + & \left[ C_F^2  \left( \frac{9}{2 \epsilon^2} - \frac{3}{4 \epsilon}
  \right) + C_F C_A  \left( \frac{11}{2 \epsilon^2} - \frac{97}{12 \epsilon}
  \right) + C_F N_f  \left( - \frac{1}{\epsilon^2} + \frac{5}{6 \epsilon}
  \right) \right] \left( \frac{\alpha_s}{4 \pi} \right)^2 \nonumber\\
  & + & \left[ C_F^3  \left( - \frac{9}{2 \epsilon^3} + \frac{9}{4
  \epsilon^2} - \frac{43}{2 \epsilon} \right) + C_F^2 C_A  \left( -
  \frac{33}{2 \epsilon^3} + \frac{313}{12 \epsilon^2} + \frac{43}{4 \epsilon}
  \right) \right. \nonumber\\
  &  & \quad + C_F C_A^2  \left( - \frac{121}{9 \epsilon^3} + \frac{1679}{54
  \epsilon^2} - \frac{11413}{324 \epsilon} \right) + C_F^2 N_f  \left(
  \frac{3}{\epsilon^3} - \frac{29}{6 \epsilon^2} + \frac{1}{\epsilon}  \left(
  \frac{23}{3} - 8 \zeta_3 \right) \right) \nonumber\\
  &  & \quad + C_F C_A N_f  \left( \frac{44}{9 \epsilon^3} - \frac{242}{27
  \epsilon^2} + \frac{1}{\epsilon}  \left( \frac{278}{81} + 8 \zeta_3 \right)
  \right) \nonumber\\
  &  & \left. \quad + C_F N_f^2  \left( - \frac{4}{9 \epsilon^3} +
  \frac{10}{27 \epsilon^2} + \frac{35}{81 \epsilon} \right) \right] \left(
  \frac{\alpha_s}{4} \right)^3 +\mathcal{O} (\alpha_s^4) \,,
\end{eqnarray}
\begin{eqnarray}
  \gamma^y (\alpha_s (\mu))  &=& \frac{d \ln y (\alpha_S (\mu))}{d \ln \mu}  =
  \sum_{n = 0} \left( \frac{\alpha_s}{4 \pi} \right)^{n + 1} \gamma^y_n \,, \nonumber \\
  \gamma^y_0 & = & - 6 C_F \,, \nonumber\\
  \gamma^y_1 & = & - C_F \left( 3 C_F + \frac{97}{3} C_A - \frac{20}{3} N_f
  T_F \right), \nonumber\\
  \gamma^y_2 & = & - C_F \Big\{ 129 C_F^2 - \frac{129}{2} C_A C_F +
  \frac{11413}{54} C_A^2 + (48 \zeta_3 - 46) C_F N_f \nonumber\\
  &&- \left( \frac{556}{27} + 48 \zeta_3 \right) C_A N_f -
  \frac{70}{27} N_f^2 \Big\}  \,.\nonumber
\end{eqnarray}
Here we briefly comment on 
the situation where the overall Yukawa coupling is renormalized in on-shell (OS) scheme.
Numerically, between these two schemes, the NNLO partial widths 
differ only by higher-order (N$^3 \tmop{LO}$ here) corrections,
while $K$ factors, instead, are quite different. The conversion
from one scheme into the other can be performed through relation \cite{hep-ph/9708255} %\cite{hep-ph/9912319}
\begin{eqnarray}
  \frac{y^{\tmop{MS}}}{y^{\tmop{OS}}} & = & 1 + \left( \frac{\alpha_s^{(N_l +
  1)} (\mu)}{\pi} \right) d_1 + \left( \frac{\alpha_s^{(N_l + 1)} (\mu)}{\pi}
  \right)^2 d_2 +\mathcal{O} (\alpha_s^3) \,,\\
  d_1 (m, \mu) & = & - C_F  \left( 1 + \frac{3}{4} L \right) \,,\nonumber\\
  d_2 (m, \mu) & = & C_F^2  \left[ \frac{7}{128} - \frac{3}{4} \zeta_3 + 3
  \zeta_2 \ln 2 - \frac{15}{8} \zeta_2 + \frac{21}{32} L + \frac{9}{32} L^2
  \right] \nonumber\\
  &  & + C_A C_F  \left[ - \frac{1111}{384} + \frac{3}{8} \zeta_3 +
  \frac{1}{2} \zeta_2 - \frac{3}{2} \zeta_2 \ln 2 - \frac{185}{96} L -
  \frac{11}{32} L^2 \right] \nonumber\\
  &  & + C_F T_F N_l  \left[ \frac{71}{96} + \frac{1}{2} \zeta_2 +
  \frac{13}{24} L + \frac{1}{8} L^2 \right] + C_F T_F  \left[ \frac{143}{96} -
  \zeta_2 + \frac{13}{24} L + \frac{1}{8} L^2 \right] \,,\nonumber
\end{eqnarray}
with $L = \ln (\mu^2 / m^2)$. $N_l = 5$ is the number of
light quarks.

\section{Ingredients of the factorization formula}
\label{app::factorization}
\subsection{The jet function}

The light quark jet function in SCET was introduced in \cite{hep-ph/0109045}, 
can be defined in terms of QCD fields as \cite{hep-ph/0603140}
\begin{equation}
  J_q (p^2) = \frac{1}{\pi N_c} \mathrm{Im} \left[ \frac{i}{\bar{n} \cdot p} 
  \int \mathrm{d}^d xe^{- ip \cdot x} \right. \left. \times \left\langle 0
  \left| \mathbf{T}  \mathrm{Tr} \left[ \frac{\slashed{\bar{n}}}{4} W^{\dagger}
  (0) \psi (0) \bar{\psi} (x) W (x) \right] \right| 0 \right\rangle \right]
\end{equation}
where $\mathbf{T}$ is the time-ordering operator, 
the trace is over color and spinor indices,
\[ W (x) = P \exp \left[ ig_s \int_{- \infty}^0 \mathrm{d} s \bar{n} \cdot A
   (x + s \bar{n}) \right] \]
denotes a $n$-collinear Wilson line, with
$n^{\mu}$ being the lightlike jet direction, 
and $\bar{n}$ being the lightlike vector subjected to $\bar{n}\cdot n =2$.

The renormalized one-loop quark jet function in momentum space reads
\begin{eqnarray}
  J_q (p^2, \mu) & = & \frac{1}{\mu^2} \left\{ \delta (x) + C_F a_s \left[ 4
  \left[ \frac{\ln x}{x} \right]_+ - 3 \left[ \frac{1}{x} \right]_+ + (7 -
  \pi^2) \delta (x) \right] +\mathcal{O} (a_s^2) \right\}, x \equiv
  \frac{p^2}{\mu^2} . \nonumber
\end{eqnarray}
However, it is often more convenient to transform the renormalized jet
function in the Laplace space
\begin{eqnarray}
  \tilde{j}^q \left( L \equiv \ln \frac{1}{e^{\gamma_E} \kappa \mu^2}, \mu \right) &
  \equiv & \int d p_J^2 e^{- \kappa p_J^2} J_q (p_J^2, \mu^2)  \,.
\end{eqnarray}
The quark jet function is known up to three-loop \cite{hep-ph/0312109,
hep-ph/0402094, hep-ph/0603140, 1008.1936, 1804.09722}. Here we show the
results up to two loops due to limited space,
\begin{eqnarray}
  \tilde{j}^q (L, \mu) & = & \sum_{n = 0} \left( \frac{\alpha_s (\mu)}{4 \pi}
  \right)^n \tilde{j}^{q (n)} (L) \,,\\
  \tilde{j}^{q (0)} & = & 1 \,,\nonumber\\
  \tilde{j}^{q (1)} & = & C_F \left( 2 L^2 - 3 L - \frac{2 \pi^2}{3} + 7
  \right) \,,\nonumber\\
  \tilde{j}^{q (2)} & = & C_A C_F \bigg\{ - \frac{22 L^3}{9} + \left(
  \frac{367}{18} - \frac{2 \pi^2}{3} \right) L^2 + L \left( 40 \zeta_3 +
  \frac{11 \pi^2}{9} - \frac{3155}{54} \right) - 18 \zeta_3 \\\nonumber
  && - \frac{155 \pi^2}{36}  - \frac{37 \pi^4}{180} + \frac{53129}{648} \bigg\} 
  +C_F^2 \bigg\{
  2 L^4 - 6 L^3 + \left( \frac{37}{2} - \frac{4 \pi^2}{3} \right) L^2 \\\nonumber
  && + L \left( - 24 \zeta_3 + 4 \pi^2 - \frac{45}{2} \right) 
   - 6 \zeta_3 + \frac{61 \pi^4}{90} + \frac{205}{8} - \frac{97 \pi^2}{12} \bigg\} \nonumber\\
  &&+C_F N_f
  \bigg\{ \frac{4 L^3}{9} - \frac{29 L^2}{9} + \left( \frac{247}{27} - \frac{2
  \pi^2}{9} \right) L + \frac{13 \pi^2}{18} - \frac{4057}{324} \bigg\}
  \,.\nonumber
\end{eqnarray}

\subsection{The soft function}
The heavy-to-light soft functions for top quark decay and charged
Higgs decay, denoted by $S_t$ and $S_H$ respectively, are defined in
Eq.~(\ref{eq:soft-def}) as
\begin{eqnarray}
  S_t(\omega) & = & \frac{1}{N_c} \sum_X {\rm Tr}
    \langle 0 | 
    \bar{Y}_v^{\dagger} (0) Y_n (0) |X\rangle
    \langle X | %\mathbf{T}
    Y_n^{\dagger} (0) \bar{Y}_v (0) |0\rangle
    \delta (\omega - n \cdot p_{\!_X}) \,,\nonumber\\
  S_H(\omega)& = & \frac{1}{N_c} \sum_X {\rm Tr}
    \langle 0 | Y_v^{\dagger} (0) Y_n (0) |X\rangle
   \langle X | %\mathbf{T}
     Y_n^{\dagger} (0) Y_v (0) |0\rangle
  \delta (\omega - n \cdot p_{\!_X})  \,.
\end{eqnarray}
The soft Wilson lines are defined as
\begin{eqnarray}
  \bar{Y}_v (x) & \equiv & \mathbf{P} \exp \left[ i g_s \int_{- \infty}^0 v
  \cdot A (x + s v) d s \right] \,,\nonumber\\
  Y_v (x) & \equiv & \overline{\mathbf{P}} \exp \left[ - i g_s \int_0^{\infty}
  v \cdot A (x + s v) d s \right] \,, \\
  Y_n^{\dagger} (x) & \equiv & \mathbf{P} \exp \left[ i g_s \int_0^{\infty} n
  \cdot A (x + s n) d s \right] \,, \nonumber
\end{eqnarray}
where $g_s$ is the strong coupling constant,
$A_\mu(x) := A_\mu^a(x)T^a$ is the (ultra)soft gluon field in the SCET,
$v^\mu \equiv p^\mu_t / m_t$ is the 4-velocity of the top quark in the rest frame
of the parent particle,
$\mathbf{P}$ ($\overline{\mathbf{P}}$)  is the (anti-)path ordering operator.
Note that the renormalized soft functions $S_t$ and $S_H$ are related to each other by
\begin{eqnarray}
  S_H (\omega, \mu) & = & \frac{m_t}{m_H} S_t \left( \omega \rightarrow
  \frac{m_t}{m_H} \omega, \mu \right) \,. 
\end{eqnarray}
They have been calculated up to three-loop order
\cite{hep-ph/0512208, 1911.04494}. The one-loop results are given by

\begin{eqnarray}
  S_t (\omega, \mu) & = & \frac{1}{\mu} \left[ \delta (x) + a_s C_F \left( -
  \zeta_2 \delta (x) - 4 \left[ \frac{1}{x} \right]_+ - 8 \left[ \frac{\ln
  x}{x} \right]_+ \right) \right], x = x_t = \frac{\omega}{\mu} \,,\nonumber\\
S_H (\omega, \mu) & =& \frac{m_t}{m_H}S_t (\frac{m_t}{m_H}\omega, \mu)
\,. \nonumber
\end{eqnarray}

\subsection{The hard function}

The Wilson coefficient renormalized in $\overline{\rm MS}$ scheme at NLO QCD for
both charged Higgs decay and top quark decay, is given by
\begin{eqnarray}
  C_{q Q}^{(\tmop{RN})} (m_t, m_H, \mu)
  & = & \frac{y^{\tmop{MS}}
          (\mu)}{y^{\tmop{LO}} (\mu_H)} \bigg\{ 1 + a_s C_F
     \bigg[ - \frac{1}{2} \ln^2
  \frac{m_t^2}{\mu^2} + \left( 2 \ln \frac{m_t^2}{m_t^2 - m_H^2} - \frac{1}{2}
  \right) \ln \frac{m_t^2}{\mu^2} \nonumber \\ 
   & & - \ln^2 (1 - \lambda) + 2 \frac{1}{\lambda} \ln (1 -
  \lambda) + 2 \tmop{Li}_2 \left( \frac{\lambda}{\lambda - 1} \right) -
  \frac{\pi^2}{12} \bigg]\bigg\}\,,\nonumber\\
  \lambda & \equiv & \frac{m_H^2}{m_t^2} + i 0^+ \,.
  \nonumber
\end{eqnarray}
For top quark decay, the Wilson coefficient is real. For charged Higgs decay,
the physical branching is chosen by
\begin{eqnarray}
  m_t^2 & \rightarrow & m_t^2 - i 0^+ \nonumber \,.
\end{eqnarray}
The corresponding hard functions are given by
\begin{eqnarray}
  H_i (m_H, m_t, \mu) & = & | C_i (m_H, m_t, \mu) |^2, i = H, t \nonumber \,.
\end{eqnarray}

\bibliographystyle{JHEP}
%\bibliography{ch}
\bibliography{draft_charged_higgs}

\end{document}